\documentclass[journal,comsoc]{IEEEtran}
\usepackage{lineno}
\usepackage{cite}
\usepackage{amsmath,amssymb}
\usepackage{tabu}
\usepackage{array}
\usepackage{comment}
\usepackage{multicol}
\usepackage[nolist]{acronym}

\ifCLASSOPTIONcompsoc
    \usepackage[caption=false, font=normalsize, labelfont=sf, textfont=sf]{subfig}
\else
\usepackage[caption=false, font=footnotesize]{subfig}
\fi

\ifCLASSINFOpdf
\else
\fi

 \usepackage{bibentry} 
\usepackage{graphicx}
\usepackage{epstopdf}
\usepackage{color}

\makeatletter
\DeclareRobustCommand\bfseriesitshape{%
\not@math@alphabet\itshapebfseries\relax
\fontseries\bfdefault
\fontshape\itdefault
\selectfont
}
\makeatother

\DeclareTextFontCommand{\textbfit}{\bfseriesitshape}

\begin{document}
\begin{acronym}

\acro{5G-NR}{5G New Radio}
\acro{3GPP}{3rd Generation Partnership Project}
\acro{AC}{address coding}
\acro{ACF}{autocorrelation function}
\acro{ACR}{autocorrelation receiver}
\acro{ADC}{analog-to-digital converter}
\acrodef{aic}[AIC]{Analog-to-Information Converter}     
\acro{AIC}[AIC]{Akaike information criterion}
\acro{aric}[ARIC]{asymmetric restricted isometry constant}
\acro{arip}[ARIP]{asymmetric restricted isometry property}

\acro{ARQ}{automatic repeat request}
\acro{AUB}{asymptotic union bound}
\acrodef{awgn}[AWGN]{Additive White Gaussian Noise}     
\acro{AWGN}{additive white Gaussian noise}

\acro{APSK}[PSK]{asymmetric PSK} 

\acro{waric}[AWRICs]{asymmetric weak restricted isometry constants}
\acro{warip}[AWRIP]{asymmetric weak restricted isometry property}
\acro{BCH}{Bose, Chaudhuri, and Hocquenghem}        
\acro{BCHC}[BCHSC]{BCH based source coding}
\acro{BEP}{bit error probability}
\acro{BFC}{block fading channel}
\acro{BG}[BG]{Bernoulli-Gaussian}
\acro{BGG}{Bernoulli-Generalized Gaussian}
\acro{BPAM}{binary pulse amplitude modulation}
\acro{BPDN}{Basis Pursuit Denoising}
\acro{BPPM}{binary pulse position modulation}
\acro{BPSK}{binary phase shift keying}
\acro{BPZF}{bandpass zonal filter}
\acro{BU}[BU]{Bernoulli-uniform}
\acro{BER}{bit error rate}
\acro{BS}{base station}
\acro{BC}{backscatter communications}
\acro{SER}{symbol error rate}
\acro{CP}{Cyclic Prefix}
\acrodef{cdf}[CDF]{cumulative distribution function}   
\acro{CDF}{cumulative distribution function}
\acrodef{c.d.f.}[CDF]{cumulative distribution function}
\acro{CCDF}{complementary cumulative distribution function}
\acrodef{ccdf}[CCDF]{complementary CDF}               
\acrodef{c.c.d.f.}[CCDF]{complementary cumulative distribution function}
\acro{CD}{cooperative diversity}

\acro{CDMA}{Code Division Multiple Access}
\acro{ch.f.}{characteristic function}
\acro{CIR}{channel impulse response}
\acro{cosamp}[CoSaMP]{compressive sampling matching pursuit}
\acro{CR}{cognitive radio}
\acro{cs}[CS]{compressed sensing}                   
\acrodef{cscapital}[CS]{Compressed sensing} 
\acrodef{CS}[CS]{compressed sensing}
\acro{CSI}{channel state information}
\acro{CCSDS}{consultative committee for space data systems}
\acro{CC}{convolutional coding}
\acro{Covid19}[COVID-19]{Coronavirus disease}
\acro{SSBC}{spectrum sharing backscatter communications}
\acro{CW}{continuous wave}

\acro{DAA}{detect and avoid}
\acro{DAB}{digital audio broadcasting}
\acro{DCT}{discrete cosine transform}
\acro{dft}[DFT]{discrete Fourier transform}
\acro{DR}{distortion-rate}
\acro{DS}{direct sequence}
\acro{DS-SS}{direct-sequence spread-spectrum}
\acro{DTR}{differential transmitted-reference}
\acro{DVB-H}{digital video broadcasting\,--\,handheld}
\acro{DVB-T}{digital video broadcasting\,--\,terrestrial}
\acro{DL}{downlink}
\acro{DSSS}{Direct Sequence Spread Spectrum}
\acro{DFT-s-OFDM}{Discrete Fourier Transform-spread-Orthogonal Frequency Division Multiplexing}
\acro{DAS}{distributed antenna system}
\acro{DNA}{Deoxyribonucleic Acid}

\acro{EC}{European Commission}
\acro{EED}[EED]{exact eigenvalues distribution}
\acro{EIRP}{Equivalent Isotropically Radiated Power}
\acro{ELP}{equivalent low-pass}
\acro{eMBB}{enhanced mobile broadband}
\acro{EMF}{electric and magnetic fields}
\acro{EU}{European union}

\acro{FC}[FC]{fusion center}
\acro{FCC}{Federal Communications Commission}
\acro{FEC}{forward error correction}
\acro{FFT}{fast Fourier transform}
\acro{FH}{frequency-hopping}
\acro{FH-SS}{frequency-hopping spread-spectrum}
\acrodef{FS}{Frame synchronization}
\acro{FSsmall}[FS]{frame synchronization}  
\acro{FDMA}{Frequency Division Multiple Access}

\acro{GA}{Gaussian approximation}
\acro{GF}{Galois field }
\acro{GG}{Generalized-Gaussian}
\acro{GIC}[GIC]{generalized information criterion}
\acro{GLRT}{generalized likelihood ratio test}
\acro{GPS}{Global Positioning System}
\acro{GMSK}{Gaussian minimum shift keying}
\acro{GSMA}{Global System for Mobile communications Association}

\acro{HAP}{high altitude platform}

\acro{IDR}{information distortion-rate}
\acro{IFFT}{inverse fast Fourier transform}
\acro{iht}[IHT]{iterative hard thresholding}
\acro{i.i.d.}{independent, identically distributed}
\acro{IoT}{Internet of Things}                      
\acro{IR}{impulse radio}
\acro{lric}[LRIC]{lower restricted isometry constant}
\acro{lrict}[LRICt]{lower restricted isometry constant threshold}
\acro{ISI}{intersymbol interference}
\acro{ITU}{International Telecommunication Union}
\acro{ICNIRP}{International Commission on Non-Ionizing Radiation Protection}
\acro{IEEE}{Institute of Electrical and Electronics Engineers}
\acro{ICES}{IEEE international committee on electromagnetic safety}
\acro{IEC}{International Electrotechnical Commission}
\acro{IARC}{International Agency on Research on Cancer}
\acro{IS-95}{Interim Standard 95}

\acro{LEO}{low earth orbit}
\acro{LF}{likelihood function}
\acro{LLF}{log-likelihood function}
\acro{LLR}{log-likelihood ratio}
\acro{LLRT}{log-likelihood ratio test}
\acro{LOS}{Line-of-Sight}
\acro{LRT}{likelihood ratio test}
\acro{wlric}[LWRIC]{lower weak restricted isometry constant}
\acro{wlrict}[LWRICt]{LWRIC threshold}
\acro{LPWAN}{low power wide area networks}
\acro{LoRaWAN}{low power long range wide area network}
\acro{NLOS}{non-line-of-sight}

\acro{MB}{multiband}
\acro{MC}{multicarrier}
\acro{MDS}{mixed distributed source}
\acro{MF}{matched filter}
\acro{m.g.f.}{moment generating function}
\acro{MI}{mutual information}
\acro{MIMO}{multiple-input multiple-output}
\acro{MISO}{multiple-input single-output}
\acrodef{maxs}[MJSO]{maximum joint support cardinality}                       
\acro{ML}[ML]{maximum likelihood}
\acro{MMSE}{minimum mean-square error}
\acro{MMV}{multiple measurement vectors}
\acrodef{MOS}{model order selection}
\acro{M-PSK}[${M}$-PSK]{$M$-ary phase shift keying}                       
\acro{M-APSK}[${M}$-PSK]{$M$-ary asymmetric PSK} 
\acro{MTC}{machine type communication}
\acro{MGF}{moment generating function} 
\acro{M-QAM}[$M$-QAM]{$M$-ary quadrature amplitude modulation}
\acro{MRC}{maximal ratio combiner}                  
\acro{maxs}[MSO]{maximum sparsity order}                                      
\acro{M2M}{machine to machine}                                                
\acro{MUI}{multi-user interference}
\acro{mMTC}{massive machine type communications}      
\acro{mm-Wave}{millimeter-wave}
\acro{MP}{mobile phone}
\acro{MPE}{maximum permissible exposure}
\acro{MAC}{media access control}
\acro{NB}{narrowband}
\acro{NBI}{narrowband interference}
\acro{NLA}{nonlinear sparse approximation}
\acro{NLOS}{Non-Line of Sight}
\acro{NTIA}{National Telecommunications and Information Administration}
\acro{NTP}{National Toxicology Program}
\acro{NHS}{National Health Service}
\acro{NB-IoT}{narrowband Internet of things}

\acro{OC}{optimum combining}                             
\acro{OC}{optimum combining}
\acro{ODE}{operational distortion-energy}
\acro{ODR}{operational distortion-rate}
\acro{OFDM}{orthogonal frequency-division multiplexing}
\acro{omp}[OMP]{orthogonal matching pursuit}
\acro{OSMP}[OSMP]{orthogonal subspace matching pursuit}
\acro{OQAM}{offset quadrature amplitude modulation}
\acro{OQPSK}{offset QPSK}
\acro{OFDMA}{Orthogonal Frequency-division Multiple Access}
\acro{OPEX}{Operating Expenditures}
\acro{OQPSK/PM}{OQPSK with phase modulation}

\acro{PAM}{pulse amplitude modulation}
\acro{PAR}{peak-to-average ratio}
\acrodef{pdf}[PDF]{probability density function}                      
\acro{PDF}{probability density function}
\acrodef{p.d.f.}[PDF]{probability distribution function}
\acro{PDP}{power dispersion profile}
\acro{PMF}{probability mass function}                             
\acrodef{p.m.f.}[PMF]{probability mass function}
\acro{PN}{pseudo-noise}
\acro{PPM}{pulse position modulation}
\acro{PRake}{Partial Rake}
\acro{PSD}{power spectral density}
\acro{PSEP}{pairwise synchronization error probability}
\acro{PSK}{phase shift keying}
\acro{PD}{power density}
\acro{8-PSK}[$8$-PSK]{$8$-phase shift keying}
\acro{PR}{primary receiver}
\acro{PT}{primary trasmitter}
 
\acro{FSK}{frequency shift keying}

\acro{QAM}{Quadrature Amplitude Modulation}
\acro{QPSK}{quadrature phase shift keying}
\acro{OQPSK/PM}{OQPSK with phase modulator }

\acro{RD}[RD]{raw data}
\acro{RDL}{"random data limit"}
\acro{ric}[RIC]{restricted isometry constant}
\acro{rict}[RICt]{restricted isometry constant threshold}
\acro{rip}[RIP]{restricted isometry property}
\acro{ROC}{receiver operating characteristic}
\acro{rq}[RQ]{Raleigh quotient}
\acro{RS}[RS]{Reed-Solomon}
\acro{RSC}[RSSC]{RS based source coding}
\acro{r.v.}{random variable}                               
\acro{R.V.}{random vector}
\acro{RMS}{root mean square}
\acro{RFR}{radiofrequency radiation}
\acro{RIS}{reconfigurable intelligent surface}
\acro{RNA}{RiboNucleic Acid}

\acro{SA}[SA-Music]{subspace-augmented MUSIC with OSMP}
\acro{SCBSES}[SCBSES]{Source Compression Based Syndrome Encoding Scheme}
\acro{SCM}{sample covariance matrix}
\acro{SEP}{symbol error probability}
\acro{SER}{symbol error rate}
\acro{SG}[SG]{sparse-land Gaussian model}
\acro{SIMO}{single-input multiple-output}
\acro{SINR}{signal-to-interference plus noise ratio}
\acro{SIR}{signal-to-interference ratio}
\acro{SISO}{single-input single-output}
\acro{SMV}{single measurement vector}
\acro{SNR}[\textrm{SNR}]{signal-to-noise ratio} 
\acro{sp}[SP]{subspace pursuit}
\acro{SS}{spread spectrum}
\acro{SW}{sync word}
\acro{SAR}{specific absorption rate}
\acro{SSB}{synchronization signal block}
\acro{SR}{secondary receiver}
\acro{ST}{secondary trasmitter}

\acro{TH}{time-hopping}
\acro{ToA}{time-of-arrival}
\acro{TR}{transmitted-reference}
\acro{TW}{Tracy-Widom}
\acro{TWDT}{TW Distribution Tail}
\acro{TCM}{trellis coded modulation}
\acro{TDD}{time-division duplexing}
\acro{TDMA}{time division multiple access}

\acro{UAV}{unmanned aerial vehicle}
\acro{uric}[URIC]{upper restricted isometry constant}
\acro{urict}[URICt]{upper restricted isometry constant threshold}
\acro{UWB}{ultrawide band}
\acro{UWBcap}[UWB]{Ultrawide band}   
\acro{URLLC}{Ultra Reliable Low Latency Communications}
         
\acro{wuric}[UWRIC]{upper weak restricted isometry constant}
\acro{wurict}[UWRICt]{UWRIC threshold}                
\acro{UE}{user equipment}
\acro{UL}{uplink}
\acro{URLLC}{ultra reliable low latency communications}

\acro{WiM}[WiM]{weigh-in-motion}
\acro{WLAN}{wireless local area network}
\acro{wm}[WM]{Wishart matrix}                               
\acroplural{wm}[WM]{Wishart matrices}
\acro{WMAN}{wireless metropolitan area network}
\acro{WPAN}{wireless personal area network}
\acro{wric}[WRIC]{weak restricted isometry constant}
\acro{wrict}[WRICt]{weak restricted isometry constant thresholds}
\acro{wrip}[WRIP]{weak restricted isometry property}
\acro{WSN}{wireless sensor network}                        
\acro{WSS}{wide-sense stationary}
\acro{WHO}{World Health Organization}
\acro{Wi-Fi}{wireless fidelity}

\acro{sss}[SpaSoSEnc]{sparse source syndrome encoding}

\acro{VLC}{visible light communication}
\acro{VPN}{virtual private network} 
\acro{RF}{radio frequency}
\acro{FSO}{free space optics}
\acro{IoST}{Internet of space things}

\acro{GSM}{Global System for Mobile Communications}
\acro{2G}{second-generation cellular networks}
\acro{3G}{third-generation cellular networks}
\acro{4G}{fourth-generation cellular networks}
\acro{5G}{5th-generation cellular networks}	
\acro{gNB}{next generation node B base station}
\acro{NR}{New Radio}
\acro{UMTS}{Universal Mobile Telecommunications Service}
\acro{LTE}{Long Term Evolution}

\acro{QoS}{Quality of Service}
\end{acronym}


%
\title{On the Performance of Spectrum Sharing Backscatter Communication Systems}

\author{Yazan H. Al-Badarneh,~\IEEEmembership{Member,~IEEE}, Ahmed Elzanaty,~\IEEEmembership{Member,~IEEE}, Mohamed-Slim Alouini,~\IEEEmembership{Fellow,~IEEE}
\thanks{Y. H. Al-Badarneh is with the department of electrical engineering, The University of Jordan, Amman, 11942 (email: yalbadarneh@ju.edu.jo).}
\thanks{A. Elzanaty and M.-S. Alouini are with the Computer, Electrical, and Mathematical Science and Engineering (CEMSE) Division, King Abdullah University of Science and Technology (KAUST), Thuwal, Makkah Province, Saudi Arabia (e-mail: \{ahmed.elzanaty, slim.alouini\}@kaust.edu.sa).}}
\markboth{}{Al-Badarneh {\MakeLowercase{\textit{et al.}}}: On the Performance of Spectrum Sharing Backscatter Communication Systems}
\maketitle

\begin{abstract} 
Spectrum sharing  backscatter communication  systems are among the most prominent technologies for ultra-low power and spectrum efficient communications. In this paper, we propose an underlay spectrum sharing backscatter communication system, in which the secondary network is a backscatter communication system. We analyze the performance of the secondary network under a transmit power adaption strategy at the  secondary transmitter, which guarantees that the interference caused by the secondary network to the primary receiver is below a predetermined threshold. We first derive a novel analytical expression for the cumulative distribution function (CDF) of the instantaneous signal-to-noise ratio of the secondary network. Capitalizing on the obtained CDF, we derive novel  accurate approximate expressions for the ergodic capacity, effective capacity and average bit error rate. We further validate our theoretical analysis using extensive Monte Carlo simulations. 

\end{abstract} 
	\acresetall 
\section{Introduction}
Legacy cellular networks have been designed with the ambition to provide voice services and high-speed data internet. Although, the race toward \ac{eMBB} has continued for  \acsu{5G} and beyond networks, other use-cases have emerged, i.e., \ac{URLLC} and \ac{mMTC} \cite{BocHeathPopovski:14,ShuppingAminSlim:20}. 
In \ac{mMTC}, a massive number of low-cost devices is required to communicate with minimal power consumption. This paradigm opens the door for several \ac{IoT} applications such as smart cities, homes, and agriculture with  21 billion expected devices by 2025 \cite{IoTstatistica:18}.

In this regard, several wireless technologies have recently emerged to achieve such low-power communications with low-cost devices. For example, \ac{LoRaWAN} and SigFoX operate in unlicensed spectrum to provide \ac{LPWAN}, while \ac{NB-IoT} technology makes use of licensed frequency bands to provide higher reliable communication \cite{ChiElz:19,RazKulSoo:17,PiaElzGioChi:17b}.  The power consumption in these networks is considerably low, and the devices have long battery life and low cost. Hence, they are suitable for many \ac{IoT} scenarios. Nevertheless,  some applications require even ultra-low-power wireless communication and self-sustainable networks, which can not be achieved with the technologies adopted for \ac{LPWAN}.

In this context, \ac{BC} is one of the most prominent technologies for ultra-low-power communications. The tag (i.e., backscatter transmitter) modulates  its information
bits over the incident electromagnetic wave and reflects it to a
receiver \cite{RezTelHer:20}.  The interest in \ac{BC} shines through its outstanding features, e.g., the transceivers in \ac{BC} do not require power-hungry components such as oscillators, mixers, and amplifiers. Therefore, they have low-cost and complexity while requiring a small amount of power that can be easily harvested from the environment\cite{VanHoaKim:18}.  Also, \ac{BC} is considered a green technology, where it usually does not require a battery. For these reasons, it is a potential candidate for self-sustainable wireless \ac{IoT} networks. 

The \ac{BC} can be classified according to the adopted architecture into three categories:  \textit{(i)  monostatic backscatter}, where a  tag modulates a signal that is generated by a reader, then reflects it back to the reader; \textit{(ii) bistatic backscatter},  where the tag backscatters the signal generated by a  \ac{CW} source to  the designated receiver;  \textit{(iii)  ambient backscatter}, where the tag backscatters an incident signal from an ambient \ac{RF} source, rather than considering a dedicated source \cite{WanGaiTel:16,LiuSmith:13}.

Besides the low power requirement,  another issue that faces wireless communications, in general, is spectrum scarcity. The enormous expansion of wireless communication systems with dedicated frequency bands (e.g., cellular networks, \ac{LPWAN}, TV broadcast, and satellite communications) results in radio spectrum congestion \cite{ChenZhaChan:20}.  This issue clearly appears in \ac{mMTC}, where we may need to allocate spectrum bands to a tremendous number of devices.
A possible solution to this problem is spectrum sharing, where multiple users can share the same frequency band under some conditions on the interference. One enabling technology for spectrum sharing is by adopting \acp{CR}, where the transceivers are aware of the environment and can accordingly adapt their transmission characteristics, e.g., the power and  frequency. In \ac{CR}, a secondary network wisely shares the spectrum with a primary network that operates on licensed frequency \cite{HossainThil:16,YazanGeoSlim:19}.

In this regard, \ac{SSBC}, which exploits both \ac{CR} and \ac{BC} technologies, has been proposed as a potential solution for both the power limitation and spectrum scarcity issues in \ac{IoT} networks. \ac{SSBC} can be classified mainly into two modes according to the interference management mechanism, i.e., \textit{(i) overlay} and \textit{(ii) underlay}.  In overlay networks, the \ac{ST} harvests energy during the operation time of the  \ac{PT}. Then, it uses such energy to transmit when the  \ac{PT}   is idle. Although \ac{SSBC} systems operating in overlay mode are self-sustainable and interference-free networks, the affordable rate for secondary networks can be limited. This can be attributed to two reasons: \textit{(i)} low harvested energy when the idle time of the \ac{PT} is significantly high;  \textit{(ii)} insufficient possible  transmission time when the \ac{PT} frequently accesses the channel, i.e., low idle time. For underlay networks, the  \ac{ST} and \ac{PT} can simultaneously access the channel. However, the  \ac{ST} should control its power to limit the interference at the \ac{PR} below a predefined threshold \cite{LeeZhangHua:13,HoaniyHan:17}.  In the following, we review the literature regarding \ac{BC} and \ac{SSBC} performance analysis.


\subsubsection{Performance Analysis of \ac{BC}}
In \cite{WanGaiTel:16}, differential modulation is proposed to reduce the \ac{BER} and maximize the sum rate in ambient \ac{BC}. The \ac{BER} of the proposed scheme is derived to demonstrate the error performance of the system. In \cite{LiuZho:17}, a coding scheme is considered for ambient \ac{BC} with tags that can have three states: no-backscatter, positive phase backscatter, and negative phase backscatter. The coding scheme maps each two consequence ternary symbols into three binary bits to increase the throughput. A higher-order modulation scheme is adopted to increase the uplink rate in \cite{BoyRoy:12}, where a coded modulation scheme with \ac{FEC} is proposed to compensate for the power loss associated with higher-order modulation. 
In \cite{JiaNiyHan:18}, a hybrid architecture that combines an ambient \ac{BC} and a classical modulator with \ac{CW} generator is presented. The modulation technique is selected to maximize the system performance, depending on the ambient environment. The throughput, energy outage, and coverage probability of the scheme are analyzed using stochastic geometry. 

A non-coherent scheme for an ambient \ac{BC} is designed in  \cite{QiaZhu:17}. In this method, the transmitter considers differential modulation. The receiver adopts either a \ac{ML} detector or an energy detector that compares the power of the recent two consecutive symbols with optimal thresholds. Also,  the \ac{BER} and outage probability are derived in closed-form. In \cite{QiaGaoZhu:17}, a semi-coherent scheme that estimates some channel-related parameters is proposed. The error performance is analyzed for ambient sources that adopt either complex Gaussian wave or \ac{PSK} modulation. 

Since many communication systems consider \ac{OFDM}, a \ac{BC} scheme is proposed in \cite{YanLiaPei:18}, where the tag modulates the ambient \ac{OFDM} wave at a rate much lower than that of the incident signal. Hence, the reflected wave can be seen as a spread spectrum signal. The optimal decoder is derived for both single and multi-antenna receivers. Then, the \ac{BER} and  data rate are investigated. 
In \cite{ElMPanSedHan:19}, the authors exploit the ambient \ac{OFDM} signal to provide $M-$ary \ac{BC}. The receiver considers a noncoherent energy detector, and the error performance is analytically investigated. 

Regarding the capacity analysis, the ergodic capacity of ambient \ac{BC} is derived in \cite{DarVerd:17}. The capacity of the ambient source that considers \ac{OFDM} is also analyzed under the interference from the \ac{BC}.   
The authors in \cite{ZhaWanAta:18} compute the constrained capacity for \ac{BPSK} and $M$-ary modulation by maximizing the mutual information between the reflected wave from the tag and received signal.

For monostatic \ac{BC} with multi-tags, an energy-efficient scheme is proposed in \cite{YazanAlouiniGeo:20}, where the tags with the largest \ac{SNR} are selected. The exact and asymptotic expressions for the ergodic and effective capacity are derived. Another tag selection scheme relying on \ac{TDMA} is suggested \cite{ZhoWanChen:17}.

\subsubsection{Spectrum Shared BC}
An \ac{RF}-powered \ac{CR} architecture is proposed in \cite{HoaNiyHan:16}. In this scheme, the  \ac{ST} can operate either in harvest-then-transmit mode or in a backscatter mode. The optimal duration of each operational mode is formulated as an optimization problem. On the other hand, the game theory is exploited for the mode selection in \cite{HoaNiyLe:17}. Another \ac{RF}-powered \ac{CR} system is proposed in \cite{KishGurDha:19}, where the  \ac{ST} performs spectrum sensing first to select the operational mode. The energy efficiency of the secondary network is maximized by optimizing the duration of various modes, e.g., energy harvesting, \ac{BC}, data transmission, and spectrum sensing phases. 

In \cite{GuoLia:19}, a cooperative  \ac{SSBC} scheme is presented, where the  \ac{PT}   simultaneously sends its own modulated symbols plus a carrier as a \ac{CW} source for the secondary \ac{BC} network. The optimal power allocation for the carrier and the optimal channel access time are computed to maximize the secondary system rate while preserving a minimum target rate for the primary network.  Another cooperative RF-powered \ac{CR} technique called symbiotic radio is visioned in \cite{LiaLi:20}. In this scheme, the primary network and the secondary \ac{BC} system are cooperative, as the primary network can benefit from the backscattered signal to enhance its channel. A possible realization of such a system is by considering the tag as a \ac{RIS}, i.e., a surface composed of many reflection elements with reconfigurable reflection coefficients \cite{RenzoSlim:19,ElzSlim:20}. In this regard, the \ac{RIS} phase profile and the beamforming are designed to minimize the transmission power while assuring a minimum \ac{QoS} for both the primary and \ac{BC} networks \cite{ZhaPoor:20}.  

\medskip

Most of the aforementioned schemes either consider the performance analysis of monostatic or mostly ambient \ac{BC}.  Each of these schemes has its own merits in terms of complexity, power efficiency, and reliability. In particular, monostatic backscatter can suffer from round-trip path loss as both the \ac{CW} source and  receiver are located on the same device, i.e., the reader. For ambient backscatter, it is the most prominent scheme from a power perspective. Nevertheless, the harvested energy and the system reliability depend on the availability of the ambient sources \cite{YanLiaPei:18,RezTelHer:20}.  On the other hand, bistatic backscatter can provide a compromise between reliability and power efficiency and a more flexible deployment of the network with separated \ac{CW} source and receiver. 


In this paper, we propose an underlay \ac{SSBC} system in which the secondary network is a \ac{BC} system consisting of a single  \ac{ST} operating as a \ac{CW} source, a single tag (T) and a single \ac{SR}. The secondary network is operating in the presence of a primary network consisting of a transmitter-receiver pair. The performance of the proposed system is analyzed under a transmit power adaption strategy at the  \ac{ST} to ensure that the interference caused by the secondary network to the \ac{PR} is below a predetermined threshold. The proposed scheme is reliable and both power and spectrum efficient, because it considers bistatic \ac{BC} and spectrum sharing. Nevertheless, its performance has not been investigated in the existing literature to the best knowledge of the authors. It is worth pointing out that the cascaded fading channel nature of the secondary system and the transmit power adaption at the \ac{ST} complicate the \ac{SNR} statistics, which in turn leads to cumbersome performance analysis.
Therefore, this paper aims to derive novel analytical expressions for the aforementioned performance metrics. In particular, we study the ergodic capacity, effective capacity, and average \ac{BER} of the secondary system. The contribution of this paper can be summarized as follows:  
\begin{itemize}
	\item We first derive a novel analytical expression for the \ac{CDF} of the instantaneous \ac{SNR} of the secondary system.
	\item Making use of the obtained CDF, we derive novel and  accurate approximate expressions for the ergodic  and effective capacities of the secondary system. 
	\item We obtain an accurate approximate expression for the average  \ac{BER} of the secondary system, which is general enough to cover a broad class of binary modulation formats suitable for \ac{BC}. 
	\item We get an approximation for the \ac{MGF}, based on which, we derive the average \ac{SER} of \ac{M-PSK} modulation.
	\item We derive insightful closed-form asymptotic expressions for the ergodic capacity, effective capacity and average \ac{BER} for large values of the average channel power gain of the secondary network. 

%
%
%
%

\end{itemize}

\medskip

The remainder of this paper is structured as follows. The system model is presented in Section II. The performance of the
proposed system is analyzed in Section III. Section IV includes the numerical results, and Section V concludes.

\section{System model}
\begin{figure}[t]\label{fig:01}
\begin{center}
\includegraphics[width=1\columnwidth]{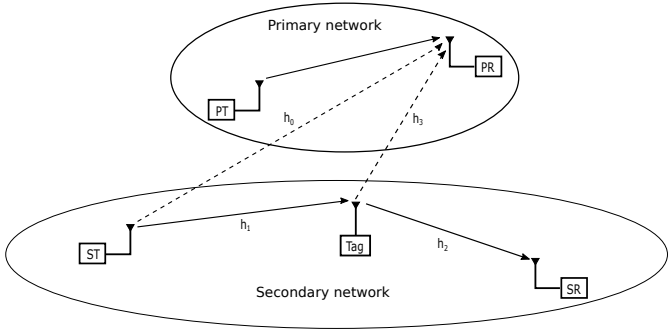}
\caption{An underlay \ac{SSBC} system in which a secondary \ac{BC} network is sharing the spectrum of a primary network.}
\end{center}
\end{figure} 

As shown in Fig. 1, we consider an underlay cognitive radio system in which a secondary network is allowed to share the spectrum of a primary network. The primary network consists of one  \ac{PT} and one \ac{PR}. The secondary network is a \ac{BC} system consisting of one  \ac{ST} that can be recognized as a \ac{CW} source, one tag (T), and one \ac{SR}. All terminals of the considered system are single antenna devices. The channel gains from the \ac{ST} to \ac{PR}  and tag are, respectively, $h_{0}$ and $h_{1}$. The channel gains from the tag to \ac{SR} and \ac{PR}  are $h_{2}$ and $h_{3}$, respectively. We assume that the aforementioned channel gains experience independent Rayleigh fading. Accordingly, the channel power gains $|h_{i}|^{2}$ for each $i \in \{0,1 ,2 , 3\}$ are exponentially distributed random variables (RVs). The \ac{CDF} and \ac{PDF} of $|h_{i}|^{2}$, denoted by $F_{|h_{i}|^{2}}(x)$ and $f_{|h_{i}|^{2}}(x)$, respectively, are given as 
\begin{equation}\label{eq:J1}
    F_{|h_{i}|^{2}}(x)= (1- e^{-\frac{x}{\lambda_{i}}}) \ u(x),
\end{equation}
\begin{equation}\label{eq:J2}
    f_{|h_{i}|^{2}}(x)=\frac{1}{\lambda_{i}} e^{-\frac{x}{\lambda_{i}}} \ u(x), 
\end{equation}
where $ i\in \{0, 1, 2, 3\}$,  $\lambda_{i}$ is the average (mean) of the RV $|h_{i}|^{2}$ and $u(x)$ is the unit step function. It should be noted that $\lambda_{i}$ is a positive real number that captures the propagation distance and the fading environment. We assume that the PT is distant from the \ac{SR} and tag; hence, the interference caused by the PT to the secondary network is negligible. Furthermore, it is assumed that the \ac{ST} is communicating with the \ac{SR} through the tag and there is no direct link between the \ac{ST} and \ac{SR}. Accordingly, the received signal at the \ac{SR} can be written as
\begin{equation}\label{J3}	
y_=  \sqrt{P}\, h_{1}\, h_{2}\, s\,  b + w, \ \  
\end{equation}
 where $P$ is the transmit power of the \ac{ST}, $s$ is the transmit signal of the \ac{ST} with $E\left[|s|^2 \right]=1$, $b$ is the information signal of the tag, and $w$ is a zero mean circularly symmetric additive white Gaussian noise (AWGN) with  variance $\sigma^{2}$. Accordingly, the  instantaneous \ac{SNR} at the \ac{SR}, denoted by $\Upsilon$, can be expressed as  
\begin{equation}\label{eq:J4}
  \Upsilon= \frac{ P|h_{1}|^{2} |h_{2}|^{2}}{\sigma^{2} }.  
  \end{equation}
It is assumed that all channel state information (CSI) is available at the \ac{ST} (i.e., $h_{i}$, for $i=1, 2, 3, 4$). Based on this CSI,  the \ac{ST} will appropriately adapt its transmit power such that the instantaneous interference power caused by the secondary network to the \ac{PR}  does not exceed a threshold $Q$, which is the maximum tolerable interference level at the \ac{PR}. Noting that the interference power at the \ac{PR}  is the sum of the interference power of the \ac{ST} $\to$ \ac{PR}  and \ac{ST} $\to$ T $\to$ \ac{PR}  links, the following condition must be satisfied   
 \begin{equation}\label{eq:J5}
  P |h_{1}|^{2} |h_{3}|^{2}+ P |h_{0}|^{2}\leq Q. 
  \end{equation} 
  
In order to maximize the instantaneous  SNR of the secondary network while maintaining the interference power at the \ac{PR}   below the threshold $Q$, the \ac{ST} will adjust its transmit power level as 
\begin{equation}\label{eq:J6}
  P=\frac{Q}{ |h_{1}|^{2} |h_{3}|^{2}+|h_{0}|^{2}}. 
  \end{equation}
Accordingly,  $\Upsilon$ in (\ref{eq:J4}) can be rewritten as  
\begin{equation}\label{eq:J7}
 \Upsilon= \frac{Q}{\sigma^{2} } \frac{ |h_{1}|^{2} |h_{2}|^{2} }{|h_{1}|^{2}|h_{3}|^{2}+ |h_{0}|^{2}}. 
  \end{equation}

%
%


\section{Performance Analysis}
In this section, we provide the performance analysis of the proposed system. We first obtain the statistics (i.e., \ac{CDF}) of the \ac{SNR} at the \ac{SR}. Then, we analyze the capacity and the average \ac{BER} based on the obtained \ac{CDF}. 

\subsection{SNR Statistics} 
In what follows, we obtain the \ac{CDF} of the instantaneous SNR $\Upsilon$ in (\ref{eq:J7}). The obtained \ac{CDF} is a key quantity of interest that is  used later in this section to analyze the capacity and average \ac{BER} of the secondary system. The \ac{CDF} of $\Upsilon$ is given by Proposition 1 below. \\

\noindent{\textbf{Proposition 1:}} The \ac{CDF} of the RV $\Upsilon$, denoted by $F_{\Upsilon}(\gamma)$, is given by 
\begin{equation}\label{eq:J8}
    F_{\Upsilon}(\gamma)= \frac{  \lambda_{0}\, \gamma\, E_{1} \bigg( \frac{ \lambda_{0}\, \gamma}{ \Omega \,\lambda_{1} \lambda_{2}}  \bigg) \,e^{ \frac{ \lambda_{0} \gamma}{\Omega \,\lambda_{1}\, \lambda_{2}}} + \lambda_{1} \lambda_{3} \gamma }{\Omega \, \lambda_{1} \lambda_{2}+  \lambda_{1} \lambda_{3}\, \gamma }, \ \  \gamma \geq0, 
    \end{equation}
where $\Omega \triangleq {Q}/{\sigma^{2}}$ and $E_{1}(\cdot )$ is the exponential integral function \cite [Eq. (5.1.4)]{abramowitz1964handbook}.

\noindent \textit{Proof:} See Appendix A. 

\subsection{Capacity Analysis}

Now, we analyze the capacity of the considered secondary system. In particular, we analyze the outage capacity, ergodic capacity and effective capacity. The channel capacity $C$ is defined  as the largest transmission rate  at which information can be transmitted with arbitrarily small probability of error \cite{proakis2008digital}.  Mathematically, the instantaneous capacity of the secondary system is given by 
\begin{equation}\label{eq:J9}
C=\log_{2} (1+\Upsilon)  \quad [\text{bits/s/Hz}],  
  \end{equation}
where  $\Upsilon$ is the instantaneous SNR in (\ref{eq:J7}). Here, the instantaneous capacity of the secondary system indicates the maximum transmission rate at a specefic time instant depending on the realization of the channel.

%
%
%


%
\subsubsection{Outage Capacity}

Since $\Upsilon$ is a RV, then $C$ is also a RV. Therefore, $C$ can drop below the transmission rate of the secondary system, $R$, leading to a communication outage. The probability of such an event is given by 
\begin{equation}\label{eq:J10}
\begin{split}
P_{out}(R)&=\Pr \{\log_{2} (1+\Upsilon) \leq R \} \\
&=\Pr \{ \Upsilon \leq \gamma_{\mathrm{th}} \} \\
&=  F_{\Upsilon}(\gamma_{\mathrm{th}}), 
\end{split} 
\end{equation}
where $\gamma_{\mathrm{th}}= 2^{R}-1$ and $F_{\Upsilon} (\cdot)$ is as given in (\ref{eq:J8}). 

In order to limit the outage probability below a certain threshold $\epsilon$, the maximum transmission rate of the system should be below a given rate indicated by the $\epsilon$-outage capacity. The $\epsilon$-outage capacity, $C_{\epsilon}$, is defined as the highest transmission rate such that outage probability $ P_{out}(R)$ is less than $\epsilon$, which can be expressed as \cite{tse2005fundamentals}
\begin{equation}\label{eq:J11}
C_{\epsilon}=\log_{2} (1+ F_{\Upsilon}^{-1}(\epsilon) ) \quad [\text{bits/s/Hz}], 
\end{equation}
where  $F_{\Upsilon}^{-1}(x) =\text{inf} \{y:F_{\Upsilon}(y) \geq x\}$.

\subsubsection{Ergodic Capacity}
We focus now on analyzing the ergodic capacity of the considered secondary system. The ergodic capacity, denoted by $\overline{C}$, can be derived by utilizing the \ac{CDF} of the RV $\Upsilon$ obtained in  (\ref{eq:J8}) as
\begin{equation}\label{eq:J12}
\begin{split}
\overline{C}=& E \left[ \log_{2} \left(1+\Upsilon \right) \right]\\
=&  \int_{0}^{\infty}  \log_{2} \left( 1+ \gamma \right) \mathrm{d} F_{\Upsilon} (\gamma)\\
=& \frac{1}{\ln(2)}  \int_{0}^{\infty} \frac{ 1-F_{\Upsilon}(\gamma)}{1+\gamma} \, \mathrm{d} \gamma. 
\end{split}
\end{equation}
Unfortunately, $\overline{C}$ above does not admit an exact analytical expression due to the exponential integral function $E_{1}(\cdot )$ of  $F_{\Upsilon}(\gamma)$ in (\ref{eq:J8}). Nevertheless, a closed-form approximation for $\overline{C}$ can be obtained by adopting the following approximation for $E_{1}(\cdot )$ \cite{6530829} 
\begin{gather}\label{eq:J13}
E_{1}(x)=4 \sqrt{2} \pi\, a_{N}\, a_{I}\, \sum_{n=1}^{N+1} \sum_{i=1}^{I+1} \sqrt{b_{n}} e^{-4 b_{n}b_{i} x},
\end{gather}
where $a_{N}= \frac{1}{2N+2}$, $a_{I}= \frac{1}{2I+2}$, $b_{n}=\frac{ \cot \left (\theta_{n-1} \right)-\cot \left (\theta_{n} \right)}{ \pi (N+1)^{-1}}$, $b_{i}=\frac{ \cot \left (\theta_{i-1} \right)-\cot \left (\theta_{i} \right)}{ \pi (I+1)^{-1}}$, $\theta_{0}=0$,  $\theta_{n}=\frac{ \pi n}{2N+2}$, $\theta_{i}=\frac{ \pi i}{2I+2}$. 
As a result, the $F_{\Upsilon} (\gamma)$ in  (\ref{eq:J8}) can be accurately approximated as 
\begin{equation}\label{eq:J14}
\begin{split}
    F_{\Upsilon}(\gamma) \approx & \frac{   \lambda_{3} \gamma}{\Omega \lambda_{2}+   \lambda_{3}\gamma }+ 4 \sqrt{2} \pi a_{N} a_{I} \sum_{n=1}^{N+1} \sum_{i=1}^{I+1} \frac{ \sqrt{b_{n}} \lambda_{0} \gamma e^{ - \frac{ \zeta_{i,n} \lambda_{0} \gamma}{\Omega \lambda_{1} \lambda_{2}}}}{\Omega\lambda_{1} \lambda_{2}+  \lambda_{1} \lambda_{3} \gamma}, 
\end{split}
\end{equation}
where $\gamma \geq0$ and $ \zeta_{i,n}\triangleq 4 b_{n} b_{i} -1 >0$.  Making use of this approximation, the ergodic capacity of the secondary system can be derived as in Proposition 2 below.  \\

\noindent{\textbf{Proposition 2:}} The ergodic capacity of the secondary system can be approximated as 
 \begin{gather}\label{eq:J15}
\begin{split}
\overline{C} \approx &  \frac{\Omega \lambda_{2}  \log_{2}  \left( \frac{\Omega \lambda_{2}}{\lambda_{3}}   \right) } {\Omega \lambda_{2}-\lambda_{3}} + \frac{4 \sqrt{2} \pi  a_{N} a_{I} \lambda_{0}}{ \ln(2)} \sum_{n=1}^{N+1} \sum_{i=1}^{I+1} \sqrt{b_{n}} \\
& \times    \left[  \frac{- \Omega \lambda_{2} \ e^{ \frac{ \lambda_{0} \zeta_{i,n}}{ \lambda_{1} \lambda_{3}}}  E_{1}\left( \frac{ \lambda_{0} \zeta_{i,n}}{ \lambda_{1} \lambda_{3}}\right) + \lambda_{3} \ e^{ \frac{ \lambda_{0} \zeta_{i,n}}{ \Omega \lambda_{1} \lambda_{2}}}  E_{1}\left( \frac{ \lambda_{0} \zeta_{i,n}}{ \Omega \lambda_{1} \lambda_{2}}\right) }{\lambda_{1} \lambda_{3} \left( \Omega \lambda_{2}- \lambda_{3} \right) }  \right],    
\end{split}
 \end{gather}
where $\lambda_{3} \neq \Omega\, \lambda_{2}$. Furthermore, the ergodic capacity of the secondary system  as a special case when $\lambda_{3} = \Omega\, \lambda_{2}$ can be approximated as 
 \begin{gather}\label{eq:J16}
\begin{split}
\overline{C_{}} \approx   &  \frac{ 1}{\ln(2)}-   \frac{4 \sqrt{2} \pi  a_{N} a_{I} \lambda_{0}}{\Omega \lambda_{1} \lambda_{2} \ln(2)} \sum_{n=1}^{N+1} \sum_{i=1}^{I+1} \sqrt{b_{n}} \\
 &\times  \left[  \ e^{ \frac{ \lambda_{0} \zeta_{i,n}}{ \Omega \lambda_{1} \lambda_{2} }}  E_{1}\left( \frac{ \lambda_{0} \zeta_{i,n}}{ \Omega \lambda_{1} \lambda_{2}} \right)  +\frac{ \lambda_{0} \ \zeta_{i,n} \ e^{ \frac{ \lambda_{0} \zeta_{i,n}}{ \Omega \lambda_{1} \lambda_{2} }}  E_{1}\left( \frac{ \lambda_{0} \zeta_{i,n}}{ \Omega \lambda_{1} \lambda_{2} }\right)  }{\Omega \lambda_{1} \lambda_{2}  } -1  \right]. 
\end{split}
 \end{gather}

\noindent \textit{Proof:} See Appendix B.

In order to gain more insight, we analyze the ergodic capacity and derive a simplified expression for it as $\lambda_{1}\to \infty$ as in Corollary 1 below.

\noindent{\textbf{Corollary 1:}}  The ergodic capacity of  the secondary system as $\lambda_{1}\to \infty$ is given by 
\begin{equation} \label{eq:J16A}
\begin{split}
\overline{C} =
\begin{cases}
\frac{\Omega \lambda_{2}  \log_{2}  \left( \frac{\Omega \lambda_{2}}{\lambda_{3}}   \right) } {\Omega \lambda_{2}-\lambda_{3}} , \ \ \ \  \lambda_{3} \neq \Omega \lambda_{2}  \\ 
\frac{ 1}{\ln(2)},  \ \ \ \ \ \ \ \ \ \ \ \ \ \ \lambda_{3} = \Omega \lambda_{2}
\end{cases}. 
\end{split}
\end{equation}

\noindent \textit{Proof:} From (\ref{eq:J8}), one can show that 
\begin{align}
& \lim\limits_{\lambda_{1}\to \infty} F_{\Upsilon}(\gamma)= \frac {\lambda_{3} \gamma }{\Omega \lambda_{2}+  \lambda_{3} \gamma }.
\label{eq:amypCDF}
\end{align}
Then, it follows immediately from the proof of Proposition 2 that the ergodic capacity is as given in  (\ref{eq:J16A}). 
 
It is worth pointing out that the result of Corollary 1 above suggests that the ergodic capacity of the secondary system initially increases with the increase of $\lambda_{1}$ and then exhibits a saturation effect for large values of $\lambda_{1}$. 

\subsubsection{Effective Capacity}
%
%
%
%
%
%
%
%
%
%
%
%
%
%
%
%

Another metric of interest is the effective capacity, which is particularly essential to quantify the system performance for delay sensitive applications \cite{1210731}. More precisely,  the effective capacity, subject a to block fading channel model, is the maximum constant arrival rate that a given service process can support in order to guarantee a statistical \ac{QoS} requirement, specified by the  \ac{QoS}  exponent $\theta$. Mathematically, the effective capacity is formulated as
\begin{eqnarray} \label{eq:J17}
\Psi (\theta) =-\frac{1}{\theta T} \log_{2} \left( E\left[ e^{-\theta T C}\right] \right), \ \theta> 0, \quad [\text{bit/s/Hz}],
\end{eqnarray}
where $C$ is a RV that represents the instantaneous capacity over a single block of length $T$ and $\theta$ is the delay  \ac{QoS}  exponent  \cite{4595458}.  It is worth mentioning that the effective capacity coincides with the ergodic capacity when the  \ac{QoS}  exponent $\theta$ equals zero. The effective capacity  of the considered secondary system can be derived in terms of the \ac{CDF} of $\Upsilon$ in  (\ref{eq:J8}) as 
\begin{gather} \label{eq:J18}
\begin{split}
\Psi (\theta) =-\frac{1}{A} \log_{2} \left( 1- A \int_{0}^{\infty} \frac{1-  F_{\Upsilon}(\gamma)} { \left(1+  \gamma \right)^{A+1}}\, \, \mathrm{d} \gamma \right)\quad [\text{bits/s/Hz}],  
\end{split}
\end{gather}
where $A \triangleq { \theta T }/{\ln(2)}$ \cite{8006945,8386679}. 

Note that a closed-form expression for integral in (\ref{eq:J18}) does not exist. Therefore, instead of the exact analysis, we use the approximation of $E_{1}(\cdot )$ in  (\ref{eq:J13}) to obtain a closed-form approximation for the effective capacity as in Proposition 3 below.

\noindent{\textbf{Proposition 3:}} The effective capacity of the secondary system can be expressed as 
\begin{equation}\label{eq:J19}
\begin{split}
\Psi (\theta)\approx&-\frac{1}{A} \log_{2} \left( 1-  \frac{ \Omega \lambda_{2} A \ {}_2F_1\left( A+1 ; 1; A+2 ; 1- \frac{\Omega  \lambda_{2}}  {\lambda_{3}}\right) }{\lambda_{3} (A+1)}  \right.\\ 
&\left. + 4  \sqrt{2} \pi  a_{N} a_{I} \lambda_{0} A \sum_{n=1}^{N+1} \sum_{i=1}^{I+1} \frac{ \sqrt{b_{n}} c_{0} E_{1} \left( \frac{ \lambda_{0}  \zeta_{i,n}}{  \lambda_{1} \lambda_{3}}\right)  e^{ \frac{ \lambda_{0} \zeta_{i,n}}{ \lambda_{1} \lambda_{3}}}}{ \left ( \Omega \lambda_{2} - \lambda_{3} \right)^{A+1}}  \right.\\ 
&\left.   + 4  \sqrt{2} \pi  a_{N} a_{I} \lambda_{0} A  \sum_{n=1}^{N+1} \sum_{i=1}^{I+1}   \sum_{k=1}^{A+1} \frac{ \sqrt{b_{n}}  c_{k}   \left( \frac{\lambda_{0} \zeta_{i,n} }{\Omega \lambda_{1} \lambda_{2}} \right)^{k-1}  }{ \left ( \Omega \lambda_{2} - \lambda_{3} \right)^{A+2-k}}  \right.\\ 
&\left.  \vphantom{ \frac{A \Omega \lambda_{2} \ {}_2F_1\left( A+1 ; 1; A+2 ; 1- \frac{\Omega  \lambda_{2}}  {\lambda_{3}}\right) }{\lambda_{3} (A+1)}  } \ \ \ \ \ \ \ \ \  \times    \Gamma \left(-k+1, \frac{ \lambda_{0}  \zeta_{i,n}}{\Omega \lambda_{1} \lambda_{2}} \right)  e^{ \frac{ \lambda_{0}  \zeta_{i,n}}{\Omega \lambda_{1} \lambda_{2}}} \right), 
\end{split}
\end{equation}
where  $\lambda_{3} \neq \Omega \lambda_{2}$, $c_{k}$, $k=0,1,..., A+1$ are constants and $A$ is an integer, ${}_2F_1\left( x,y; z;w\right)$ is the Gauss hypergeometric function \cite[Eq. (9.34.7)]{jeffrey2007table} and $\Gamma(\cdot,\cdot)$ is the upper incomplete Gamma function \cite[Eq.8.350.2)]{jeffrey2007table}.  Furthermore, the effective capacity of the secondary system as a special case for $\lambda_{3} = \Omega \lambda_{2}$ can be approximated as 
\begin{gather}\label{eq:J20} 
\begin{split}
\Psi (\theta)\approx&-\frac{1}{A} \log_{2} \left( 1-  \frac{A}{A+1}   + \frac{4 \sqrt{2} \pi a_{N} a_{I} \lambda_{0} A}{ \Omega \lambda_{1} \lambda_{2}}\right.\\ 
&\left. \times \sum_{n=1}^{N+1} \sum_{i=1}^{I+1}    \sqrt{b_{n}} \textit{U}\left( 2;1-A; \frac{ \zeta_{i,n} \lambda_{0} }{  \Omega \lambda_{1} \lambda_{2}}\right) \right),  
\end{split}
\end{gather}
for an  arbitrary $A$, where $\textit{U}\left( a;b;z \right)=\frac{1}{\Gamma(a)} \int_{0}^{\infty} e^{-zt} t^{a-1} (1+t)^{b-a-1}dt$, $a > 0$ is the Tricomi hypergeometric function \cite[Eq. (13.1.3)]{abramowitz1964handbook}.

\noindent \textit{Proof:} See Appendix C. 

In Corollary 2 below, we quantify effective capacity of the secondary system as  $\lambda_{1}\to \infty$. 

\noindent{\textbf{Corollary 2:}}  The effective capacity of  the secondary system as $\lambda_{1}\to \infty$ is given by 

\begin{gather} \label{eq:J20A}
\Psi (\theta) =-  \frac{1}{A} \log_{2} \left( 1-  \frac{ \Omega \,\lambda_{2}\, A \  {}_2F_1\left( A+1 ; 1; A+2 ; 1- \frac{\Omega  \lambda_{2}}  {\lambda_{3}}\right) }{\lambda_{3} (A+1)} \right),
\end{gather}
for arbitrary values of $A$, $\Omega$, $\lambda_{2}$, and $\lambda_{3}$. 

\noindent \textit{Proof:} Considering \eqref{eq:amypCDF},
Corollary 2 follows immediately from the proof of Proposition 3.



\subsection{Error Performance Analysis}
Here, we analyze the error performance of  the secondary system in terms of the \ac{BER} for various binary modulation formats and the \ac{SER} for \ac{M-PSK}.
\subsubsection{Average \ac{BER} Analysis}
We  consider  the average \ac{BER} for several binary modulation schemes for which the conditional \ac{BER}, $P_{e}(p,q)$, is characterized by \cite{1096424}
\begin{gather}\label{eq:J21}
P_{e}(p,q)= \frac{ \Gamma(p, q \Upsilon) }{ 2\, \Gamma(p)},
\end{gather}
where $\Upsilon$ is the instantaneous SNR of the secondary system as in  (\ref{eq:J7}). The parameters $p$ and $q$ are positive real constants that refer to  a specific binary modulation format.  It should be noted that  (\ref{eq:J21}) is general enough to cover many binary modulation formats, such as the binary phase shift keying (BPSK), binary frequency shift keying (BFSK), noncoherent binary frequency shift keying (NBFSK) and differential binary phase shift keying (DBPSK). Please refer to \cite{5957242} for more details. 

 The average \ac{BER} of the secondary system, denoted by $\overline {P_{e}}$, can be expressed as 
\begin{gather}\label{eq:J22}
\begin{split}
\overline{ P_{e}}(p,q)=&  \int_{0}^{\infty} \frac{ \Gamma(p, q \gamma) }{ 2 \Gamma(p)} \, \mathrm{d}F_{\Upsilon}(\gamma)\\
\end{split}
\end{gather}
In Proposition 4 below, we derive an approximate expression for the average \ac{BER} of the secondary system. \\

\noindent{\textbf{Proposition 4:}} The average \ac{BER} of the secondary system can be approximated as   
\begin{equation}\label{eq:J23}
\begin{split}
\overline{ P_{e}}(p,q)\approx & \frac{p}{2}\left(\frac{q \Omega \lambda_{2}}{ \lambda_{3}} \right)^p e^{\frac{ q \Omega \lambda_{2}} {\lambda_{3}}}  \Gamma \left(-p, \frac{ q \Omega \lambda_{2}} {\lambda_{3} } \right)  \\
+ &4 \sqrt{2} \pi a_{N} a_{I} \sum_{n=1}^{N+1} \sum_{i=1}^{I+1}  
 \frac{\sqrt{b_{n}}  p \lambda_{0}}{2 \lambda_{1} \lambda_{3}}\left(\frac{q \Omega \lambda_{2}}{ \lambda_{3}} \right)^p e^{\frac{ q \Omega \lambda_{2}} {\lambda_{3}}+ \frac{ \lambda_{0} \zeta_{i,n}}{\lambda_{1} \lambda_{3}}} \\
& \times \Gamma \left(-p, \frac{ q \Omega \lambda_{2}} {\lambda_{3} } +\frac{ \lambda_{0} \zeta_{i,n}}{\lambda_{1} \lambda_{3}}\right).
\end{split}
\end{equation}
\noindent \textit{Proof:} See Appendix D. 

In Corollary 3 below, we quantify the average \ac{BER} of the secondary system as  $\lambda_{1}\to \infty$. 

\noindent{\textbf{Corollary 3:}}  The average \ac{BER}  of  the secondary system as $\lambda_{1}\to \infty$ is given by 
\begin{equation} \label{eq:J23A}
\overline{ P_{e}}(p,q)\approx \frac{p}{2}\left(\frac{q \Omega \lambda_{2}}{ \lambda_{3}} \right)^p e^{\frac{ q \Omega \lambda_{2}} {\lambda_{3}}}  \Gamma \left(-p, \frac{ q \Omega \lambda_{2}} {\lambda_{3} } \right).
\end{equation}

\noindent \textit{Proof:} Corollary 3 follows immediately after making use of  \eqref{eq:amypCDF} in the proof of Proposition 4.


\subsubsection{Average {SER} Analysis}
In this subsection, we analyze the average \ac{SER} of the secondary system with \ac{M-PSK} modulation format  adopting \ac{MGF} based approach. We first focus on the \ac{MGF} of the instantaneous \ac{SNR}, $\Upsilon$, which is defined as 
\begin{equation} \label{eq:J24}
\mathcal{M}_{\Upsilon}(s)=E \left[ e^{-s \Upsilon}\right]=  \int_{0}^{\infty} e^{-s \gamma} \,  \mathrm{d}F_{\Upsilon}(\gamma). 
\end{equation}
Based on $F_{\Upsilon}(\gamma)$ obtained earlier in (\ref{eq:J14}), we provide an approximate expression for the \ac{MGF} of the instantaneous \ac{SNR} in Corollary 4 below. 

\noindent{\textbf{Corollary 4:}}  The \ac{MGF} of  the instantaneous \ac{SNR}, $\Upsilon$, can be approximated as 
\begin{equation} \label{eq:J25}
\begin{split}
\mathcal{M}_{\Upsilon}(s) \approx & \frac{ \Omega \lambda_{2}s}{ \lambda_{3}}  e^{\frac{  \Omega \lambda_{2}s} {\lambda_{3}}}  \Gamma \left(-1, \frac{ \Omega \lambda_{2}s} {\lambda_{3} } \right)  \\
+ &4 \sqrt{2} \pi a_{N} a_{I} \sum_{n=1}^{N+1} \sum_{i=1}^{I+1}  
 \frac{\sqrt{b_{n}} \lambda_{0}\Omega \lambda_{2}s}{ \lambda_{1} \lambda_{3}^{2}}  e^{\frac{ \Omega \lambda_{2}s} {\lambda_{3}}+ \frac{ \lambda_{0} \zeta_{i,n}}{\lambda_{1} \lambda_{3}}} \\
& \times \Gamma \left(-1, \frac{ \Omega \lambda_{2}s} {\lambda_{3} } +\frac{ \lambda_{0} \zeta_{i,n}}{\lambda_{1} \lambda_{3}}\right).
\end{split}
\end{equation}
\noindent \textit{Proof:} One can easily show that setting  $p=1$ and $q=s$ in (\ref{eq:J22}), yields
\begin{equation} \label{eq:J26}
\overline{ P_{e}}(1,s)=  \int_{0}^{\infty} \frac{1}{2} e^{-s \gamma} \, \mathrm{d}F_{\Upsilon}(\gamma)=\frac{1}{2}\mathcal{M}_{\Upsilon}(s). 
\end{equation}
This implies that $\mathcal{M}_{\Upsilon}(s)=2\overline{ P_{e}}(1,s)$, where $\overline{ P_{e}}(1,s)$ can be obtained by setting  $p=1$ and $q=s$ in (\ref{eq:J23}). 

In view of the \ac{MGF} derived above, the average \ac{SER} of \ac{M-PSK} can be evaluated as \cite{SimonAlouini:05}
\begin{equation} \label{eq:J27}
\overline{ P_{s}}=\frac{1}{\pi} \int_{0}^{\pi-\frac{\pi}{M}} \mathcal{M}_{\Upsilon} \left(\frac{\alpha_{M}}{\sin^{2}(\theta)}\right) \mathrm{d}\theta,  
\end{equation}
where $\alpha_{M}= \sin^{2}(\pi /M)$.

\section{ Numerical Results}
In this section, we use the derived analytical results to numerically evaluate the performance of the secondary system. The accuracy of the analytical expressions is confirmed using Monte Carlo simulations, where the performance is being evaluated by averaging $10^{6}$ independent realizations. For simplicity but without loss of generality, the noise power is assumed to be normalized to unity (i.e., $\sigma^{2}=1$) in this section. 

\begin{figure}[h]\label{fig:02}
\begin{center}
\includegraphics[width=1\columnwidth,clip]{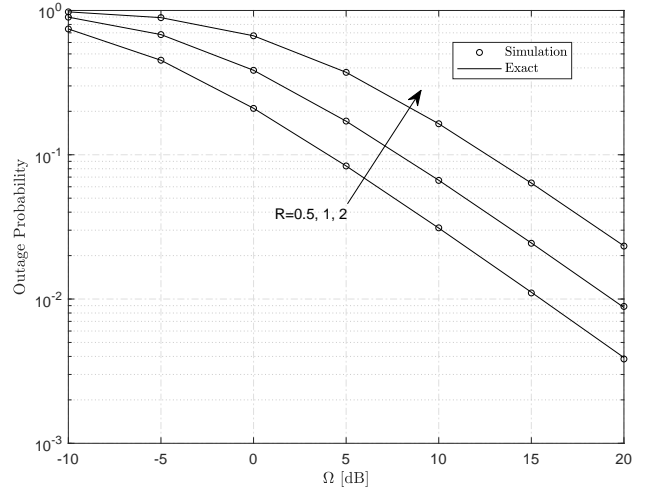}
\caption{  Outage probability versus peak interference to noise power ratio $\Omega$ for  $\lambda_{1}=4$,  $\lambda_{2}=3$,  $\lambda_{0}=\lambda_{3}=1$, and   $R=0.5, 1, 2$.   }
\end{center}
\end{figure} 

In Fig. 2, the outage probability is plotted against the peak interference to noise power ratio $\Omega=Q/\sigma^{2}$ for $\lambda_{1}=4$,  $\lambda_{2}=3$,  $\lambda_{0}=\lambda_{3}=1$, and  $R \in \{0.5, 1, 2\}$. It can be readily observed that the outage probability decreases as $\Omega$ increases. This is rather intuitive, since the \ac{PR}  tolerates higher interference power as $Q$ increases. Accordingly, the \ac{ST} can adjust its transmit power to a higher level, which results in a lower outage probability. 

\begin{figure}[h]\label{fig:02}
\begin{center}
\includegraphics[width=1\columnwidth,clip]{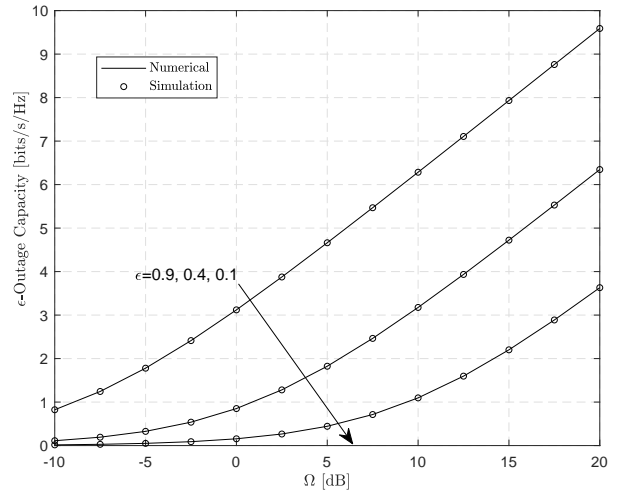}
\caption{ Outage capacity versus peak interference to noise power ratio $\Omega$ for $\lambda_{1}=2$, $\lambda_{2}=3$, $\lambda_{0}=\lambda_{3}=1$, and $\epsilon=0.1, 0.4, 0.9$.  }
\end{center}
\end{figure} 

\begin{figure}[h]\label{fig:02}
\begin{center}
\includegraphics[width=1\columnwidth]{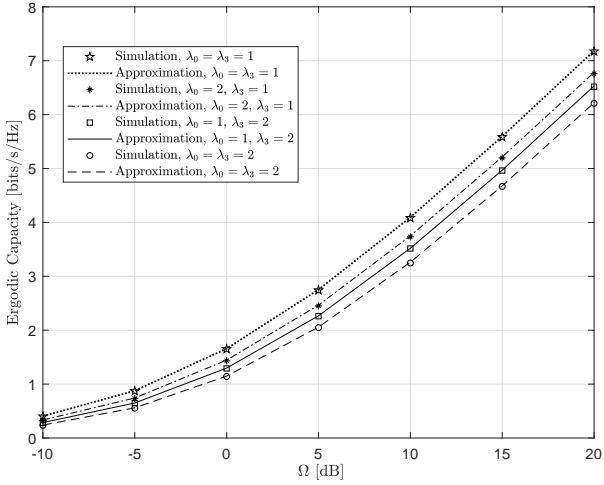}
\caption{   Ergodic capacity versus peak interference to noise power ratio $\Omega$ for $\lambda_{1}=4$, $\lambda_{2}=3$ and  different values of $\lambda_{0}$ and  $\lambda_{3}$.   }
\end{center}
\end{figure} 
In Fig. 3, the $\epsilon$-outage capacity is depicted against $\Omega$ for $\lambda_{1}=2$, $\lambda_{2}=3$, $\lambda_{0}=\lambda_{3}=1$, and $\epsilon \in \{0.1, 0.4, 0.9\}$. As illustrated,  the $\epsilon$-outage capacity of the secondary system improves as $\Omega$ increases.

Fig. 4, illustrates the ergodic capacity as a function of $\Omega$ for $\lambda_{1}=4$, $\lambda_{2}=3$, and  different values of $\lambda_{0}$ and  $\lambda_{3}$. The analytical curves obtained via the approximations for ergodic capacity obtained earlier in (\ref{eq:J15}) and (\ref{eq:J16}) with  $I=N=14$. We note that the approximations for ergodic capacity match perfectly with the Monte Carlo simulations. As expected, the ergodic capacity increases as $\Omega$ increases. However, the ergodic capacity decreases as $\lambda_{0}$ or $\lambda_{3}$ increases. This is due to the fact that as $\lambda_{0}$ or $\lambda_{3}$ increases, the  interference power caused by the secondary network to the \ac{PR}  increases, hence, the \ac{ST} will decrease its transmit power to satisfy the interference constraint at the \ac{PR}, which consequently decreases the ergodic capacity of the secondary system.   

\begin{figure}[h]\label{fig:02}
\begin{center}
\includegraphics[width=1\columnwidth]{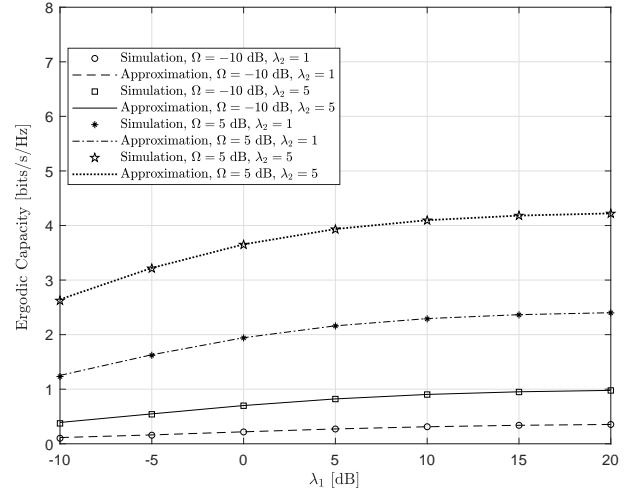}
\caption{   Ergodic capacity versus $\lambda_{1}$  for $\lambda_{0}=0.1$, $\lambda_{3}=1$ and  different values of $\Omega$ and  $\lambda_{2}$.   }
\end{center}
\end{figure} 

In Fig. 5, we investigate the ergodic capacity as a function of $\lambda_{1}$ for $\lambda_{0}=0.1$, $\lambda_{3}=1$, and different values of $\Omega$ and  $\lambda_{2}$. The analytical curves obtained via the approximations in (\ref{eq:J15}) and (\ref{eq:J16}). We note that the approximations for the ergodic capacity coincide with the Monte Carlo simulations. Most importantly, we observe that the ergodic capacity initially increases with the increase of $\lambda_{1}$ and then saturates for large vaues of $\lambda_{1}$. In addition, we show that the anayltical and simulation curves for the ergodic capacity approach that obtained in (\ref{eq:J16A}) as $\lambda_{1} \to \infty$.  

In Fig. 6, we plot the effective capacity as a function of delay  \ac{QoS}  exponent $A$ for $\lambda_{1}=2$, $\lambda_{2}=3$, $\lambda_{0}=\lambda_{3}=1$, and different values of $\Omega$. We evaluate the effective capacity using the numerical integration for arbitrary values of $A$ and the analyitcal expressions obtained via the approximations in (\ref{eq:J20}) and (\ref{eq:J21}) for an integer $A$ with $I=N=14$. As demonstrated in the figure, the simulation results confirm the accuracy of the curves obtained through the numerical integration and the approximat expressions. 

In Fig. 7, we plot the average \ac{BER}  versus $\Omega$ for various binary modulation formats for $\lambda_{1}=4$, $\lambda_{2}=3$, $\lambda_{0}=0.1$ and $\lambda_{3}=1$. It is evident from the figure that the obtained expression in (\ref{eq:J23}) provides a highly accurate approximation of the \ac{BER} performance of the seconday system.  Finally, we confirmed the correctness of our theoretical analysis, for various system parameters, through  simulations.

\begin{figure}[h]\label{fig:02}
\begin{center}
\includegraphics[width=1\columnwidth]{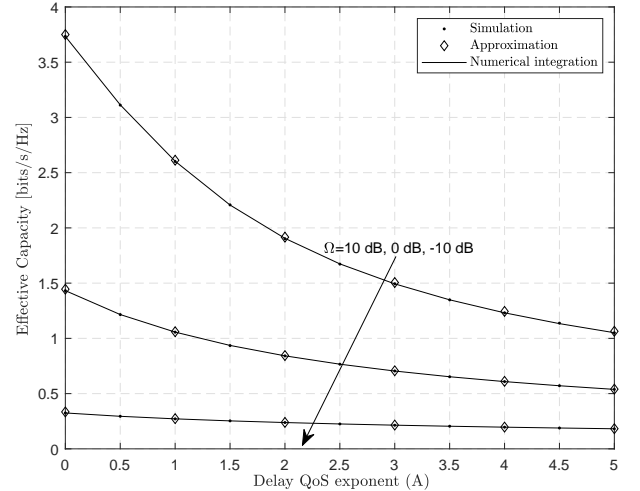}
\caption{   Effective capacity versus delay  \ac{QoS}  exponent $A$ for $\lambda_{1}=2$,  $\lambda_{2}=3$,  $\lambda_{0}=\lambda_{3}=1$, and  $\Omega=-10, 0, 10$ dB. }
\end{center}
\end{figure}

\begin{figure}[h]\label{fig:02}
\begin{center}
\includegraphics[width=1\columnwidth,clip]{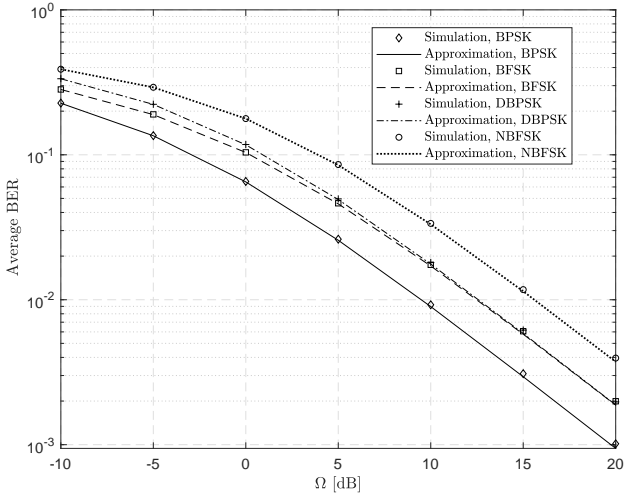}
\caption{   Average \ac{BER} of various binary modulation formats versus interference to noise power ratio $\Omega$ for $\lambda_{1}=4$,  $\lambda_{2}=3$, $\lambda_{0}=0.1$, $\lambda_{3}=1$.} 
\end{center}
\end{figure}

\section{Conclusion} 
We proposed an underlay SSBC system in which a secondary BC system shared the spectrum of a primary network. The performance of the secondary BC system is analyzed employing a transmit power adaption strategy at the \ac{ST} that satisfies the instantaneous interference constraint at the \ac{PR}.  We derived an exact analytical expression for the \ac{CDF} of the SNR of the secondary system, based on which, we studied the ergodic capacity, effective capacity and average BER. More specifically, we derive novel and  accurate analytical approximations for the ergodic and effective capacities, together with the average BER. We further obtained insightful closed-form asymptotic expressions for the aforementioned performance metrics as $\lambda_{1} \to \infty$.  For instance, it was evident from the asymptotic ergodic capacity that the performance will initially improve as $\lambda_{1}$ increases and then saturate as $\lambda_{1}$ grows large. Finally, we confirmed the correctness of our theoretical analysis, for various system parameters, through Monte Carlo simulations. 

\section*{Appendix A \\ Proof Of Proposition 1}
The \ac{CDF} of the RV $\Upsilon$ in (\ref{eq:J7}) can be expressed as 
\begin{eqnarray} \label{eq:Q1}
F_{\Upsilon} (\gamma)=\Pr \{ \Upsilon \leq \gamma\}=\Pr \bigg \{  \frac{ \Omega |h_{1}|^{2} |h_{2}|^{2} }{|h_{1}|^{2}|h_{3}|^{2}+ |h_{0}|^{2}} \leq \gamma\bigg \},  
\end{eqnarray}
where $\Omega= {Q}/{\sigma^{2}}$.  In veiw of $|h_{i}|^{2}$, $ i\in \{0, 1, 2, 3\}$, being independent exponentially distributed RVs with the \ac{CDF} and \ac{PDF} given in (\ref{eq:J1}) and  (\ref{eq:J2}), respectively,  $F_{\Upsilon} (\gamma) $ can be obtained as shown in (\ref{eq:Q2}) at the bottom of the page.  Utilizing \cite[Eq. (3.353.5)]{jeffrey2007table} to evaluate $I_{1}$ of  (\ref{eq:Q2}) below, $F_{\Upsilon} (\gamma) $ can be finally expressed as in (\ref{eq:J8}) after some basic algebraic manipulations.
\begin{figure*}[h]
\hrule
\begin{align} \label{eq:Q2}
\begin{split}
F_{\Upsilon} (\gamma)&= \int_{0}^{\infty}  \int_{0}^{\infty}  \int_{0}^{\infty} \Pr \bigg \{   |h_{2}|^{2}  \leq \frac{ \gamma ( x_{1} x_{3}+ x_{0})}{\Omega   x_{1}}\bigg \} f_{|h_{0}|^{2}} (x_{0})   f_{|h_{1}|^{2}} (x_{1})  f_{|h_{3}|^{2}} (x_{3}) \ \mathrm{d}x_{0} \ \mathrm{d}x_{1} \ \mathrm{d}x_{3} \\
&= \int_{0}^{\infty}  \int_{0}^{\infty}  \int_{0}^{\infty}  \bigg(1- e^{-\frac{\gamma ( x_{1} x_{3}+ x_{0})}{\Omega \lambda_{2} x_{1}}} \bigg)     \frac{ 1}{\lambda_{0}} e^{-\frac{x_{0}}{ \lambda_{0}} } \  \frac{ 1}{\lambda_{1}} e^{-\frac{x_{1}}{ \lambda_{1}} } \ \frac{ 1}{\lambda_{3}} e^{-\frac{x_{3}}{ \lambda_{3}} } \  \mathrm{d}x_{0} \ \mathrm{d}x_{1} \ \mathrm{d}x_{3} \\
&= 1-    \int_{0}^{\infty}  \int_{0}^{\infty}  \int_{0}^{\infty}  e^{-\frac{\gamma ( x_{1} x_{3}+ x_{0})}{\Omega \lambda_{2} x_{1}}} \ \frac{ 1}{\lambda_{0}} e^{-\frac{x_{0}}{ \lambda_{0}} } \  \frac{ 1}{\lambda_{1}} e^{-\frac{x_{1}}{ \lambda_{1}} } \ \frac{ 1}{\lambda_{3}} e^{-\frac{x_{3}}{ \lambda_{3}} } \  \mathrm{d}x_{0} \ \mathrm{d}x_{1} \ \mathrm{d}x_{3} \\
&= 1-    \int_{0}^{\infty}  \int_{0}^{\infty} \ \bigg ( \int_{0}^{\infty}   e^{-\frac{\gamma  x_{0}}{\Omega \lambda_{2} x_{1}}} \  \frac{ 1}{\lambda_{0}}  e^{-\frac{x_{0}}{ \lambda_{0}} } \  \mathrm{d}x_{0} \bigg) \   e^{- \frac{\gamma x_{3} }{ \Omega \lambda_{2} } }     \  \frac{ 1}{\lambda_{1}} e^{-\frac{x_{1}}{ \lambda_{1}} } \ \frac{ 1}{\lambda_{3}} e^{-\frac{x_{3}}{ \lambda_{3}} }  \ \mathrm{d}x_{1} \ \mathrm{d}x_{3}\\
&= 1-    \int_{0}^{\infty} \frac{ \Omega \lambda_{2} x_{1} }{\Omega \lambda_{2} x_{1}+\lambda_{0} \gamma }   \bigg( \int_{0}^{\infty}  \ e^{- \frac{\gamma x_{3} }{ \Omega \lambda_{2} } }    \ \frac{ 1}{\lambda_{3}} e^{-\frac{x_{3}}{ \lambda_{3}} }    \ \mathrm{d}x_{3} \bigg)     \frac{ 1}{\lambda_{1}} e^{-\frac{x_{1}}{ \lambda_{1}} } \ \mathrm{d}x_{1}\\
&= 1-    {\frac {{\Omega}^{2}{{\lambda_{2}}}^{2}}{  \Omega {\lambda_{2}}+{ \lambda_{3}\gamma }  }} \underbrace{  \int_{0}^{\infty} {\frac {{x_{1}}}{ \Omega { \lambda_{2} x_{1}}+ \lambda_{0}\gamma }}
  \  \frac{ 1}{\lambda_{1}} e^{-\frac{x_{1}}{ \lambda_{1}} } \ \mathrm{d}x_{1}}_\text{$I_{1}$}. \\
&= 1-  {\frac {{\Omega}^{2}{{\lambda_{2}}}^{2}}{  \Omega {\lambda_{2}}+{ \lambda_{3} \gamma }  }}  \left( \frac{ - \lambda_{0} \gamma E_{1} \left( \frac{ \lambda_{0} \gamma}{ \Omega \lambda_{1} \lambda_{2}}  \right) e^{ \frac{ \lambda_{0} \gamma}{\Omega \lambda_{1} \lambda_{2}}} + \Omega \lambda_{1} \lambda_{2}}{{\Omega}^{2}{{\lambda_{2}}}^{2} \lambda_{1} } \right). 
\end{split}
\end{align}
\hrule
\end{figure*}

\
\section*{Appendix B \\ Proof Of Proposition 1}
Making use of (\ref{eq:J14}) in (\ref{eq:J12}), yields 
\begin{equation}\label{eq:Q3}
\begin{split}
\overline{C}\approx & \frac{1}{\ln(2)}  \underbrace{ \int_{0}^{\infty}  \frac{   \Omega \lambda_{2}}{ \left( \Omega \lambda_{2}+  \lambda_{3}  \gamma \right) (1+\gamma) }\, \mathrm{d} \gamma}_\text{$I_{2}$} - \frac{4 \sqrt{2} \pi a_{N} a_{I}}{ \ln(2)}   \\
 &\times \sum_{n=1}^{N+1} \sum_{i=1}^{I+1}  \underbrace{ \int_{0}^{\infty}   \frac{  \sqrt{b_{n}} \lambda_{0} \gamma e^{ - \frac{ \zeta_{i,n} \lambda_{0} \gamma}{\Omega \lambda_{1} \lambda_{2}}}}{ \left( \Omega\lambda_{1} \lambda_{2}+  \lambda_{1} \lambda_{3} \gamma \right) (1+\gamma)} \, \mathrm{d} \gamma} _\text{$I_{3}$}. 
\end{split}
\end{equation}

Using partial fraction decomposition, $I_{2}$ can be readily obtained as 
 \begin{equation}\label{eq:Q4}
\begin{split}
I_{2}&= \frac{\Omega\lambda_{2} }{{\Omega\lambda_{2}-\lambda_{3}}} \int_{0}^{\infty} \left( \frac{- \lambda_{3}}{\Omega\lambda_{2}+  \lambda_{3} \gamma} + \frac{1}{1+\gamma}\right) \, \mathrm{d} \gamma  \\
&= \frac{\Omega \lambda_{2}  \ln  \left( \frac{\Omega \lambda_{2}}{\lambda_{3}}   \right) } {\Omega \lambda_{2}-\lambda_{3}}, \ \  \ \  \Omega \neq \frac{ \lambda_{3}}{\lambda_{2}}. 
\end{split}
\end{equation}
Similarly, $I_{3}$ can be expressed as
 \begin{equation}\label{eq:Q5}
\begin{split}
 I_{3}= \frac{ \sqrt{b_{n}} \lambda_{0} }{{\Omega\lambda_{2}-\lambda_{3}}} & \int_{0}^{\infty} \left( \frac{\Omega \lambda_{2}   e^{ - \frac{ \zeta_{i,n} \lambda_{0} \gamma}{\Omega \lambda_{1} \lambda_{2}}}  }{\Omega \lambda_{1}\lambda_{2}+  \lambda_{1} \lambda_{3}\gamma}   - \frac{  e^{ - \frac{ \zeta_{i,n} \lambda_{0} \gamma}{\Omega \lambda_{1} \lambda_{2}}}  }{\lambda_{1}+\lambda_{1}\gamma}\right) \, \mathrm{d} \gamma. 
\end{split}
\end{equation}
Utilizing the integral formula \cite[Eq. (3.352.4)]{jeffrey2007table}, $I_{3}$ can be finally evaluated as 
 \begin{equation}\label{eq:Q6}
\begin{split}
 I_{3} =  \ \frac{ \sqrt{  b_{n}} \Omega \lambda_{0} \lambda_{2} \ e^{ \frac{ \lambda_{0} \zeta_{i,n}}{ \lambda_{1} \lambda_{3}}}  E_{1}\left( \frac{ \lambda_{0} \zeta_{i,n}}{ \lambda_{1} \lambda_{3}}\right)  }{\lambda_{1} \lambda_{3} \left( \Omega \lambda_{2}- \lambda_{3} \right) }   -  \frac{ \sqrt{b_{n}} \lambda_{0}  e^{ \frac{ \lambda_{0} \zeta_{i,n}}{ \Omega \lambda_{1} \lambda_{2}}}  E_{1}\left( \frac{ \lambda_{0} \zeta_{i,n}}{ \Omega \lambda_{1} \lambda_{2}}\right)  }{\lambda_{1} \left( \Omega \lambda_{2}- \lambda_{3} \right) } ,    
\end{split}
\end{equation}
where  $\lambda_{3} \neq \Omega \lambda_{2}$.  Substituting  (\ref{eq:Q4}) and (\ref{eq:Q6}) into (\ref{eq:Q3}),  after some basic algebraic manipulations, yields the desired result in  (\ref{eq:J15}). The ergodic capacity  of the secondary system for the special case  $\lambda_{3} = \Omega \lambda_{2}$ can be obtained by substituting  $\lambda_{3} = \Omega \lambda_{2}$ in (\ref{eq:Q3}). Accordingly, 
 \begin{equation}\label{eq:Q7} 
\begin{split}
\overline{C} \approx & \frac{1}{ \ln(2) } \int_{0}^{\infty} \frac{1 }{\left( 1+\gamma \right) ^{2}}  \, \mathrm{d} \gamma  \\
&-\frac{4 \sqrt{2} \pi a_{N} a_{I} \lambda_{0} }{ \Omega \lambda_{1} \lambda_{2} \ln(2)  } \sum_{n=1}^{N+1} \sum_{i=1}^{I+1}   \int_{0}^{\infty}   \frac{  \sqrt{b_{n}} \gamma e^{ - \frac{ \zeta_{i,n} \lambda_{0} \gamma}{\Omega \lambda_{1} \lambda_{2}}}}{ (1+\gamma)^2} \, \mathrm{d} \gamma  \\
=&\frac{1}{\ln(2)} -\frac{4 \sqrt{2} \pi a_{N} a_{I} \lambda_{0} }{\Omega \lambda_{1} \lambda_{2} \ln(2)  } \sum_{n=1}^{N+1} \sum_{i=1}^{I+1} \left ( \int_{0}^{\infty} \frac{ \sqrt{b_{n}} e^{ - \frac{ \zeta_{i,n} \lambda_{0} \gamma}{ \Omega \lambda_{1} \lambda_{2}}}} {1+\gamma} \right.\\ 
&\left.  \ \ \ \ \ \ \ \ \   \ \ \ \ \ \ \ \ \   \ \ \ \ \ \ \ \ \   \ \ \ \ \ \ \ \ \  - \int_{0}^{\infty} \frac{ \sqrt{b_{n}} e^{ - \frac{ \zeta_{i,n} \lambda_{0} \gamma}{ \Omega \lambda_{1} \lambda_{2}}}} {(1+\gamma)^{2}} \right). 
\end{split}
\end{equation}
Utilizing the identities \cite[Eq. (3.352.4)]{jeffrey2007table} and \cite[Eq. (3.353.2)]{jeffrey2007table} to evaluate  the involved integrals above, after some algebraic manipulations, $\overline{C}$ is finally as expressed in (\ref{eq:J16}).  

\section*{Appendix C \\  Proof Of Proposition 3}
Making use of (\ref{eq:J14}) in the integral of (\ref{eq:J18}), yields 
\begin{gather}\label{eq:Q8}
\begin{split}
\int_{0}^{\infty} \frac{1-  F_{\Upsilon}(\gamma)} { \left(1+  \gamma \right)^{A+1}} \approx &  \underbrace{ \int_{0}^{\infty}  \frac{   \Omega \lambda_{2}}{ \left( \Omega \lambda_{2}+   \lambda_{3}\gamma \right) (1+\gamma)^{A+1}}\, \mathrm{d} \gamma}_\text{$\Xi_{1}$}  \\
- 4 \sqrt{2} \pi a_{N} a_{I} \sum_{n=1}^{N+1}& \sum_{i=1}^{I+1}     \underbrace{ \int_{0}^{\infty}   \frac{ \sqrt{b_{n}}  \lambda_{0} \gamma e^{ - \frac{ \zeta_{i,n} \lambda_{0} \gamma}{\Omega \lambda_{1} \lambda_{2}}}}{ \left( \Omega\lambda_{1} \lambda_{2}+  \lambda_{1} \lambda_{3}\gamma \right) (1+\gamma)^{A+1}} \, \mathrm{d} \gamma} _\text{$\Xi_{2}$}.
\end{split}
\end{gather}
 $\Xi_{1}$ can be evaluated with the help \cite[Eq. (3.227.1)]{jeffrey2007table} as 
\begin{equation}\label{eq:Q9}
\Xi_{1}=  \frac{\Omega \lambda_{2}}{\lambda_{3} (A+1)}  \ {}_2F_1\left( A+1 ; 1; A+2 ; 1- \frac{\Omega  \lambda_{2}} {\lambda_{3}}  \right).  
\end{equation}
It is not tractable to find a closed-form expression for  $\Xi_{2}$ above for an arbitrary $A$. However, it is possible to find a closed-form for  $\Xi_{2}$ when $A$ is an integer. Applying partial fraction decomposition,   $\Xi_{2}$ can be expressed as 
\begin{equation}\label{eq:Q10}
\begin{split}
\Xi_{2}=& \int_{0}^{\infty} \frac{ \sqrt{b_{n}}  \lambda_{0} c_{0} e^{ - \frac{ \zeta_{i,n} \lambda_{0} \gamma}{\Omega \lambda_{1} \lambda_{2}}}}{ \left ( \Omega \lambda_{2} - \lambda_{3} \right)^{A+1} \left(  \gamma + \frac{ \Omega \lambda_{2}}{ \lambda_{3}} \right)}  \, \mathrm{d} \gamma\\
&+ \sum_{k=1}^{A+1} \int_{0}^{\infty} \frac{ \sqrt{b_{n}}  \lambda_{0} c_{k}  e^{ - \frac{ \zeta_{i,n} \lambda_{0} \gamma}{\Omega \lambda_{1} \lambda_{2}}}}{ \left ( \Omega \lambda_{2} - \lambda_{3} \right)^{A+2-k} \left(  \gamma + 1\right)^{k}}  \, \mathrm{d} \gamma,
\end{split}
\end{equation}
where $c_{k}$, $k=0,1,..., A+1$ are constants. Utilizing \cite[Eq. (3.352.4)]{jeffrey2007table} and \cite[Eq. (3.382.4)]{jeffrey2007table}, after some algebraic manipulations, $\Xi_{2}$  can be evaluated as 
\begin{equation}\label{eq:Q11}
\begin{split}
\Xi_{2} = &  \frac{ \sqrt{b_{n}}  \lambda_{0} c_{0} E_{1} \left( \frac{ \lambda_{0}  \zeta_{i,n}}{  \lambda_{1} \lambda_{3}}\right)  e^{ \frac{ \lambda_{0} \zeta_{i,n}}{ \lambda_{1} \lambda_{3}}}}{ \left ( \Omega \lambda_{2} - \lambda_{3} \right)^{A+1}} \\
&+  \sum_{k=1}^{A+1} \frac{ \sqrt{b_{n}}  \lambda_{0} c_{k}   \left( \frac{\lambda_{0} \zeta_{i,n} }{\Omega \lambda_{1} \lambda_{2}} \right)^{k-1}  }{ \left ( \Omega \lambda_{2} - \lambda_{3} \right)^{A+2-k}} \Gamma \left(-k+1, \frac{ \lambda_{0}  \zeta_{i,n}}{\Omega \lambda_{1} \lambda_{2}} \right)  e^{ \frac{ \lambda_{0}  \zeta_{i,n}}{\Omega \lambda_{1} \lambda_{2}}}, 
\end{split}
\end{equation}
where   $\lambda_{3} \neq \Omega \lambda_{2}$ and $A$ is an integer.  Making use of  (\ref{eq:Q9}), (\ref{eq:Q11}) and (\ref{eq:Q8}) in (\ref{eq:J18})  yields the desired result in  (\ref{eq:J19}).   However, the  effective capacity  of the secondary system for the special case  when  $\lambda_{3} = \Omega \lambda_{2}$ can be obtained by substituting   $\lambda_{3} = \Omega \lambda_{2}$ in (\ref{eq:Q8}). Accordingly, 

\begin{gather}\label{eq:Q12}
\begin{split}
\int_{0}^{\infty} \frac{1-  F_{\Upsilon}(\gamma)} { \left(1+  \gamma \right)^{A+1}}\, \, \mathrm{d} \gamma=&  \int_{0}^{\infty} \frac{1} {\left( 1+\gamma \right)^{A+2}}  \, \mathrm{d} \gamma \\
-   \frac{4 \sqrt{2} \pi a_{N} a_{I} \lambda_{0}}{ \Omega \lambda_{1} \lambda_{2}} &\sum_{n=1}^{N+1} \sum_{i=1}^{I+1}  \underbrace{ \int_{0}^{\infty} \frac{ \sqrt{b_{n}} \gamma e^{ - \frac{ \zeta_{i,n} \lambda_{0} \gamma}{ \Omega \lambda_{1} \lambda_{2} }} }{\left( 1+\gamma \right)^{A+2}}  \, \mathrm{d} \gamma}_\text{$\Xi_{3}$}.
\end{split}
\end{gather}
$\Xi_{3}$ can be evaluated with the help \cite[Eq. (39)]{1576535} as 
\begin{gather}\label{eq:Q13}
\begin{split}
\Xi_{3}=  \sqrt{b_{n}} \textit{U}\left( 2;1-A; \frac{ \zeta_{i,n} \lambda_{0}}{  \Omega \lambda_{1} \lambda_{2}}\right). 
\end{split}
\end{gather}
Making use of  (\ref{eq:Q12}) and (\ref{eq:Q13}) in (\ref{eq:J18})  yields the desired result in  (\ref{eq:J20}). 


\section*{Appendix D \\ Proof Of Proposition 4}
Using integration by parts,  $\overline {P_{e}}(p,q)$ in  (\ref{eq:J22}) can be expressed as 
\begin{gather}\label{eq:Q14}
\begin{split}
\overline{ P_{e}}(p,q)= \frac{q^{p}}{ 2 \Gamma(p)}  \int_{0}^{\infty} e^{-q \gamma} \gamma^{p-1} F_{\Upsilon}(\gamma) \, \mathrm{d} \gamma.
\end{split}
\end{gather}
Making use of (\ref{eq:J14}) in (\ref{eq:Q14}), yields 
\begin{gather}\label{eq:Q15}
\begin{split}
\overline{ P_{e}}(p,q)\approx&\frac{q^{p}}{ 2 \Gamma(p)}   \int_{0}^{\infty}  \frac{  \lambda_{3}  e^{-q \gamma} \gamma^{p} }{\Omega \lambda_{2}+ \gamma \lambda_{3} } \, \mathrm{d} \gamma +\frac{4 \sqrt{2} \pi a_{N} a_{I} q^{p}}{ 2 \Gamma(p)}\\
&  \times \sum_{n=1}^{N+1} \sum_{i=1}^{I+1}  \int_{0}^{\infty}  \frac{ \sqrt{b_{n}} \lambda_{0}  e^{ - \frac{ \zeta_{i,n} \lambda_{0} \gamma}{\Omega \lambda_{1} \lambda_{2}}}  \gamma^{p} e^{-q \gamma} }{\Omega\lambda_{1} \lambda_{2}+ \gamma \lambda_{1} \lambda_{3}}   \, \mathrm{d} \gamma.
\end{split}
\end{gather}
Utilizing \cite[Eq. (3.383.10)]{jeffrey2007table} to evaluate the involved intgerals above, after basic algebric manipulations, $\overline {P_{e}}$ is finally as expressed in (\ref{eq:J23}). 

\bibliographystyle{IEEEtran}
\bibliography{yvlc}
\end{document}